\documentclass[11pt,letterpaper]{article}

\usepackage{graphicx}

\begin{document}

\def\beq{\begin{equation}}
\def\eeq{\end{equation}}
\def\bea{\begin{eqnarray}}
\def\eea{\end{eqnarray}}

\def\a{\alpha}
\def\r{\rho}
\def\s{\sigma}
\def\m{\mu}
\def\n{\nu}
\def\k{\kappa}
\def\g{\gamma}
\def\L{\Lambda}
\def\D{\Delta}
\def\la{\langle}
\def\ra{\rangle}
\def\o{\omega}
\def\d{\delta}
\def\p{\partial}
\def\Se{$S_E$ }
\def\Sa{$S_{\rm atmo}$ }

\def\tphi{\tilde{\phi}}
\def\tu{\tilde{u}}
\def\hv{\hat{v}}

\def\half{\textstyle{\frac{1}{2}}}
\def\quarter{\textstyle{\frac{1}{4}}}

\begin{center} {\LARGE \bf
Black hole entanglement entropy\\
\vskip 2mm
regularized in a freely falling frame}
\end{center}

\vskip 5mm
\begin{center} \large
{Ted Jacobson$^{*}$\footnote{E-mail: jacobson@umd.edu} and Renaud Parentani$^{\dagger}$\footnote{E-mail: Renaud.Parentani@th.u-psud.fr}}\end{center}

\vskip  0.5 cm
{\centerline{$^{*}$Department of Physics}}
{\centerline{\it University of Maryland}} 
{\centerline{College Park, MD 20742-4111, USA}} 

\vskip  0.5 cm
{\centerline{$^{\dagger}$Laboratoire de 
Physique Th\'eorique}}
{\centerline{\it CNRS UMR 8627,
Universit\'e Paris-Sud}} 
{\centerline{91405 Orsay Cedex,
France}} 

\vskip 1cm

\begin{abstract}
We compute the black hole horizon entanglement entropy \Se
for a  massless scalar field,
first with a hard cutoff
and then with high frequency dispersion,
both imposed in a frame
that falls freely across the horizon.
Using WKB methods, we find that \Se is finite for
a hard cutoff or super-luminal dispersion,
because the mode oscillations do
not diverge at the horizon and the contribution
of high transverse momenta is cut off by the
angular momentum barrier. For sub-luminal dispersion
the entropy depends
on the behavior at arbitrarily high transverse momenta.
In all cases it scales with the horizon area.
For the hard cutoff it is linear in the cutoff,
rather than quadratic. This discrepancy from the
familiar result arises from the difference between the
free-fall frame and the static frame in which a
cutoff is usually imposed.
In the super-luminal case the entropy scales
 with a fractional power of the cutoff
that depends on
the index of the dispersion relation.
Implications for the possible relation between regularized
entanglement entropy and  the Bekenstein-Hawking
entropy are discussed.
An appendix provides an explicit derivation
of the entangled, thermal nature of the near-horizon
free fall vacuum for a dispersive scalar field in four dimensions.

\end{abstract}

\newpage
\section{Introduction}

All indications suggest that the Bekenstein-Hawking entropy of a
black hole
\beq
S_{\rm BH}= {A \over 4L_p^2}
\eeq
is true thermodynamic
entropy\cite{Bekenstein:1972tm,Bekenstein:1973ur,Hawking:1976de}.
(Here $A$ is the area of the horizon
and $L_p$ the Planck length.) At
the microscopic level it must therefore correspond to the
logarithm of the number of
thermodynamically relevant
states.
One can write down a formal path integral expression for the partition
function and entropy of the quantum gravitational statistical
ensemble\cite{Gibbons:1976ue}, which in the classical approximation yields precisely
this entropy, however the microscopic nature of the states thereby
counted remains obscure.

It would appear that an unavoidable contribution to the entropy
budget is the entanglement entropy $S_E$
across the horizon of quantum
fields propagating on the black hole
background\cite{Sorkin83,Bombelli:1986rw}.
The entanglement
entropy arises from the spacelike correlations in the freely
falling vacuum (giving rise to Hawking radiation),
on account of which the restriction to the region
outside the horizon is a mixed state.
This can also be thought of as entropy of
the thermal atmosphere of the black
hole\cite{'tHooft:1984re,Zurek:1985gd},
particularly referring
to the near-horizon region where most of the entropy is localized.
It is difficult to see any rationale for omitting the
entanglement/atmosphere
entropy.
However, in the standard semiclassical evaluation it diverges
due to the infinite
density of states at
the horizon.

One can only conclude that the divergence in $S_E$
must be somehow cut off by quantum gravity effects.
A simple dimensional analysis of the geometry
suggests that when cut off at Planck scale, the
entanglement entropy across the horizon will scale as
$A/L_p^2$, just like the Bekenstein-Hawking entropy.
If so, then $S_E$ would make an important contribution to
$S_{BH}$.  This presents a ``species problem,"
since $S_E$ depends on the number and nature of quantum
fields, whereas the Bekenstein-Hawking entropy is {\it
universally} given by $A/4L_p^2$.

A natural solution to this problem may exist
however\cite{Susskind:1994sm,Jacobson:1994iw}: the value of the
Planck length $L_p=(\hbar G/c^3)^{1/2}$ appearing in $S_{BH}$
involves the low energy gravitational constant $G$, whose value
should depend, though renormalization effects,  on the quantum
matter fields. The shift of $1/G$ produced by the matter fields
could induce a shift of $S_{BH}$ that corresponds precisely to
$S_E$. The formal arguments of
Refs.~\cite{Susskind:1994sm,Jacobson:1994iw} support this
possibility, as do explicit regularized computations for free field
theory such as in~\cite{Demers:1995dq}.
 However, mismatches seem to arise
 for contributions of
  vector fields~\cite{Kabat:1995eq}
and in odd dimensional spacetimes~\cite{Kim:1996bp}.
 On the other hand, very general
 arguments~\cite{Solodukhin:1995ak,Larsen:1995ax}
 indicate
   that if one focuses not on entanglement
 entropy but only on the contribution of
 matter to the entropy derived from the
 thermal partition function for gravity plus matter
 (which may be the same thing in some cases),
 then the relation between black hole
 entropy and renormalization
 of $G$ is universal. A purely thermodynamic
 argument~\cite{Jacobson:1995ab}
   suggests the even stronger statement
 that $G$ is wholly determined by the
entropy of local Rindler horizons.

Previous studies of black hole
entanglement entropy
with a physically motivated regulator
(as opposed to a technical regulator such as
Pauli-Villars)
have adopted a cutoff scheme where
a highest momentum \cite{Bombelli:1986rw}
or shortest distance \cite{'tHooft:1984re} was
imposed in the static frame. In both cases it was found that the
entropy scales as $A/\epsilon^2$ where $\epsilon$ is the cutoff
length.
However, the
equivalence principle suggests that if there
is a cutoff on the
freely falling vacuum
imposed by quantum gravity it should be
physically determined in
a freely falling frame.

The aim of this
paper is to investigate
the nature of the regulated entanglement entropy when
a freely falling
momentum cutoff
$\L$
is imposed.
We
shall consider both a hard momentum cutoff, and a cutoff imposed
by the introduction of dispersion at high wavevectors, and we
carry out the analysis using WKB methods.
We find
that unlike
what is found in the static frame,
 the entropy scales
linearly with the cutoff
as
$A\k_v \L$
for a hard cutoff, and with a similar but more complicated scaling law
in the case of dispersion. In this expression
$\k_v$ is the surface gravity measured in the freely falling frame
in which $\L$ has been defined.
When that frame is at rest at infinity, $\k_v$ reduces to the
standard surface gravity.

A surprising conclusion may be drawn from our results. If the
freely falling cutoff $\L$ is identified with the Planck mass,
then the regulated entanglement entropy is negligibly small
compared to the Bekenstein-Hawking entropy for macroscopic black
holes. This would provide a different resolution to the species
problem mentioned above
(unless the number of species were of order $M_p/\k_v$).
On the other hand, as argued in \cite{Parentani:2000ts,Parentani:2002mb},
it is also possible
that the cutoff arises from quantum gravitational effects in the
black hole background,
and that its value equals the Planck mass when measured in
the {\it static} frame.\footnote{If quantum
horizon fluctuations
fuzz out the horizon over the much longer scale
$(R_sL_p^2)^{1/3}$, as suggested
in~\cite{Sorkin:1996sr,Casher:1996ct,Marolf:2003bb},
then the question arises whether the entropy is
cut off at this scale or instead if the cutoff ``floats"
with the fluctuations.}
Such a cutoff would
appear as $\L\sim M_p^2/\k_v$
in the free-fall
frame,
and the regulated entropy
would
be of order $A/L_p^2$ after all.
We shall return to a discussion of this issue at the end of the
paper.

Whatever the regularization scheme one chooses,
the
divergence in the entanglement entropy can be
traced to two distinct sources. First, the density of modes at
fixed frequency and angular momentum diverges due to an infinite
number of oscillations as the horizon is approached. Second, there
is the sum over all angular momenta.
In our approach,
we find that in the case of
both the hard cutoff and dispersion the WKB mode density is
regulated. In effect, the dispersion introduces a maximum radial
wave number contributing to the thermal atmosphere of the black
hole. We further find that the mode density vanishes at a finite
angular momentum in the case of a hard cutoff or super-luminal
dispersion, while it decreases to zero asymptotically at large
angular momenta in the sub-luminal case. The
angular momentum sum
is therefore finite in the former cases, while
in the latter case it
depends on the dispersion
at asymptotically large momenta and
cannot be
determined within
the WKB approximation because of the asymptotically vanishing mode
density.

The effect of sub-luminal dispersion on entanglement entropy has
been previously investigated in Ref.~\cite{Chang:2003sa}. There
the dispersion was introduced in the static (accelerated) frame of
the black hole, rather than in the freely falling frame. In that
case, there is a piling up of transplanckian modes
at some radius greater than the horizon radius,
so the mode density at fixed angular momentum remains
infinite and therefore the entropy diverges. This is not the
result reported in \cite{Chang:2003sa}, where it was instead
concluded that the entropy is finite. We believe this conclusion
was based on an incorrect evaluation of the entropy.

In following sections of this paper we first review the nature of
the divergent entanglement entropy in the usual relativistic case.
Next we introduce a hard momentum cut-off in the freely falling
frame and analyze how it regulates the divergences. Finally we
introduce dispersion at high momenta also in the freely falling
frame, and analyze its impact on the divergences of the
entanglement entropy.
A long appendix provides details on the
field modes, the entangled, thermal nature of the
free fall vacuum, and the
WKB evaluation of the mode density
governing the entanglement entropy.

\section{Thermal atmosphere of a black hole}

We consider nonrotating, static black hole spacetimes and adopt
Painlev\'{e}-Gullstrand (PG) coordinates since they are adapted to
the free-fall frame in which the momentum cutoff or dispersion
will be introduced. A free-fall frame is defined by a family of
radial infalling geodesics all with the same velocity at infinity.
The
corresponding
PG line element in four dimensions takes the form
\beq ds^2 = d\tau^2 -(d\rho-v(\rho)d\tau)^2 -r(\rho)^2 (d\theta^2 +
\sin^2\theta d\phi^2), \label{ds2} \eeq
in units with the speed of light $c=1$ (see e.g.
Ref.~\cite{Martel:2000rn}).
The spacetime has $\tau$-translation and
rotational symmetry.
The coordinate $\tau$ differs from the
Schwarzschild-like time by a constant scaling and
an additive function of $\rho$.
The function $v(\rho)$ is
the velocity $d\rho/d\tau$ of the radial geodesics, along which $\tau$
measures the proper time. These geodesics are orthogonal to the
surfaces of constant $\tau$, on which $\rho$ measures radial proper
distance. We consider the case $v(\rho)<0$, so that these
geodesics are infalling. A surface $\rho=\rho_H$ on which
$v^2(\rho_H)=1$ is lightlike and is a future event horizon, i.e.
no timelike curve can exit the interior region $\rho<\rho_H$.

Expanding
around the horizon one has
\beq v(\rho)=-1+\kappa_v(\rho-\rho_H) + \cdots. \label{kappa}\eeq
As we shall see, the non-linear terms are not relevant to the
entropy. The constant $\kappa_v$ depends on the free-fall frame at
the horizon.\footnote{A coordinate independent local expression for
$\k_v$ is $(1/2)(s\cdot\nabla\xi^2)/(s\cdot\xi)^2$, where $s$ is the
outward radial unit spacelike vector orthogonal to the free-fall
worldline at the horizon and $\xi$ is the Killing vector with
arbitrary normalization (which cancels out in this ratio).}
It is equal to the surface gravity defined with respect to the
Killing vector $\partial_\tau$, which
coincides with the standard surface gravity $\k$ of the horizon
in the case that $v(\rho)\rightarrow0$ as $\rho\rightarrow\infty$.
For spacetimes in which $v(\rho)$ approaches a nonzero constant
$v_\infty$, $\partial_\tau$ has norm $(1-v_\infty^2)^{1/2}$
at spatial infinity, so
\beq \kappa_v=\kappa \ (1-v_\infty^2)^{1/2}.\label{kvk}\eeq
For the example of a Schwarzschild black hole of mass $M$, the
frame that falls from rest at infinity has
$v(\rho)=-(2GM/\rho)^{1/2}$. In this case the surface gravity is
$\kappa=1/4GM$, and the area radius $r(\rho)$ is equal to $\rho$,
the radial proper distance
measured at constant $\tau$.\footnote{The vanishing of
$r''(\rho)$ is equivalent to the vanishing of the (ingoing or
outgoing) radial null-null component of the Ricci tensor.
The condition $r'(\rho)=1$ holds only if $v(\rho)$ vanishes at
infinity.}

When a black hole forms from collapse of matter, the quantum state
of any relativistic field tends towards the Unruh
vacuum~\cite{Unruh:1977ga,Brout:1995rd}. This
stationary state coincides with the vacuum as defined by freely
falling (FF) observers for modes with FF frequencies much larger
than the surface gravity $\k$.
In this state fluctuations on either side of the horizon are
correlated in such a way that when tracing over the inside degrees
of freedom, the reduced density matrix for the outgoing modes has
the thermal form $\rho_{\rm ext}=Z^{-1} \exp(-H/T_H)$, where $H$
is the Hamiltonian generating the Killing time translations
(normalized at infinity) and
$T_H=\kappa/2\pi$
is the Hawking
temperature~\cite{Wald:1975kc,Parker:1975jm,Hawking:1976ra}.
(Here and below we adopt units with $\hbar=c=1$.)
Some of the outgoing thermal
quanta reach spatial infinity and constitute Hawking
radiation. Most of them however fall back across the
horizon into the black hole, forming a thermal
``atmosphere"~\cite{Membrane}.
 The entropy of this atmosphere
$S_{\rm atmo}$ consists of the part of the
entanglement (von Neumann) entropy $-{\rm
Tr}(\rho_{\rm ext}\ln \rho_{\rm ext})$ contributed by
the near-horizon
modes,
that is, not including the
entropy of the
few low angular momentum Hawking quanta that escape to infinity.
\Sa is
in fact
dominated by
initially
outgoing modes
with large angular
momenta which are turned back by the angular momentum
barrier very close to the horizon.

Given the symmetries and the thermal character of
the reduced density matrix $\rho_{\rm ext}$,
the atmosphere entropy can be expressed as a {\it mode sum}
in the form
\beq S_{\rm atmo}=\int d\omega\, \sum_{\ell=0}^{\infty}(2\ell+1)\,
n(\omega,\ell)\, s_{\rm therm}(\omega/T_H), \label{Satmo}\eeq
where $\omega$ is the Killing frequency, $\ell$ is the angular
momentum, $s_{\rm therm}(\omega/T_H)$
is the thermal entropy per
mode,\footnote{Strictly
speaking there is a weak $\ell$-dependence of the mode entropy due
to the transmitted part of the infalling modes. These start from
infinity in the ground state and partially penetrate the effective
potential barrier, whereas the reflected parts of the near-horizon
outgoing modes are thermally populated. Since the transmitted part
vanishes for $\ell\gg\o/\k$, this $\ell$ dependence can be safely
neglected when computing the entropy of the atmosphere.}
and $n(\omega,\ell)$ is the density of outgoing
modes entangled to inside configurations.

The entropy diverges for two independent reasons. First, for each
fixed $\omega$ and $\ell$, the mode density has a UV divergence
logarithmic in the radial momentum cut-off. This arises from the
infinite number of mode oscillations caused by the infinite blue
shift at the horizon. Secondly, the sum over angular momenta is
unbounded, producing a UV divergence quadratic in the angular
momentum. Notice that for a 1+1 dimensional black hole only the
first, logarithmic divergence is present.

\section{Freely falling cut-off}

We are interested in regulating the atmosphere entropy by modifying
the field propagation at high momenta as defined in a freely falling
frame. This is in contrast to 't Hooft's ``brick wall"
approach~\cite{'tHooft:1984re} to regulating the entropy, where
the cut-off is defined on static time slices.
We first consider the simple case of a hard momentum cut-off.

Since the momentum cut-off is stationary and spherically
symmetric, the atmosphere entropy still has the form
(\ref{Satmo}). The entropy per mode is unchanged, but the mode
density $n(\o,\ell)$ is  cut-off.
To compute the atmosphere
entropy we therefore only need to evaluate the cut-off mode
density $n_{\rm cut}(\o,\ell)$.
The calculation is identical for any free-fall frame in which the
cutoff is imposed, so for notational simplicity in the following
we adopt the one for which $v_\infty=0$.

We consider a massless scalar field in the spacetime with line
element (\ref{ds2}).
The field theoretic justification for our
evaluation of the entropy, while essential for a full understanding,
is sequestered in the Appendix so as not to obscure the conceptual
structure and simplicity of the entropy evaluation via WKB methods.
The reader may wish to read the Appendix at this point before coming
back to the main flow of the entropy computations.

The modes of fixed frequency and angular
momentum take the form
\beq \varphi_{\omega\ell m} = \exp(-i\omega \tau )r^{-1}
u_{\omega\ell}(\rho) Y_{\ell m}(\theta,\phi).
\label{partialwaves}\eeq
 Since the relevant
modes have very short wavelength, the radial mode functions
$u_{\omega\ell}(\rho)$ are accurately described by the WKB form
$u\sim \exp(i\int k\, d\rho )$, where the radial momentum
$k=k_{\omega\ell}(\rho)$ obeys the relativistic dispersion relation
$g^{\mu\nu}k_\mu k_\nu=0$, which here takes the form
\beq (\omega-v(\rho)k)^2= k^2 + \ell(\ell+1)/r^2.
\label{reldisp}\eeq
The quantity $\Omega = \omega-vk$
is the free-fall frequency, whereas
$\omega$ is the conserved Killing frequency. We introduce the UV
cutoff by counting only modes with $k<\L$ for some fixed $\L \gg
\kappa$.
The cutoff is imposed on the radial component of momentum only,
while the transverse (angular) momentum is unrestricted. It turns
out however that the result would
differ little
if the cutoff were
imposed on the total momentum,
since the angular momentum barrier imposes an
effective transverse
momentum cut-off of order $(\o\L)^{1/2} \ll \L$.

The atmosphere entropy is defined as the entropy of the density
matrix obtained after tracing over the field degrees of freedom
localized inside the horizon. We shall
evaluate the entropy here with the help of the WKB approximation, in
which the modes
are characterized by
a well-defined momentum
$k_{\omega\ell}(\rho)$, the solution of (\ref{reldisp}),
at any radial position $\rho$. This allows us to count modes that
lie outside the horizon and below the momentum cutoff.

\begin{figure}
 \includegraphics[angle=0,width=3cm]{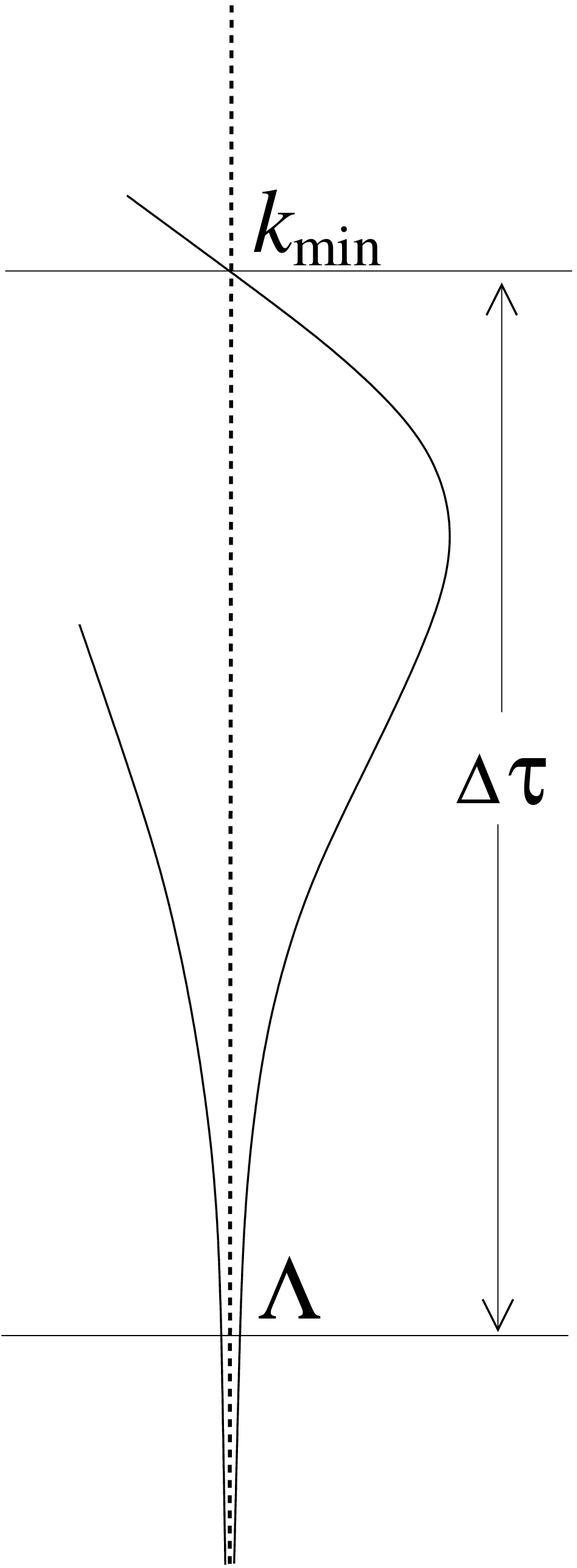}
 \hspace{1cm}
  \includegraphics[angle=0,width=3cm]{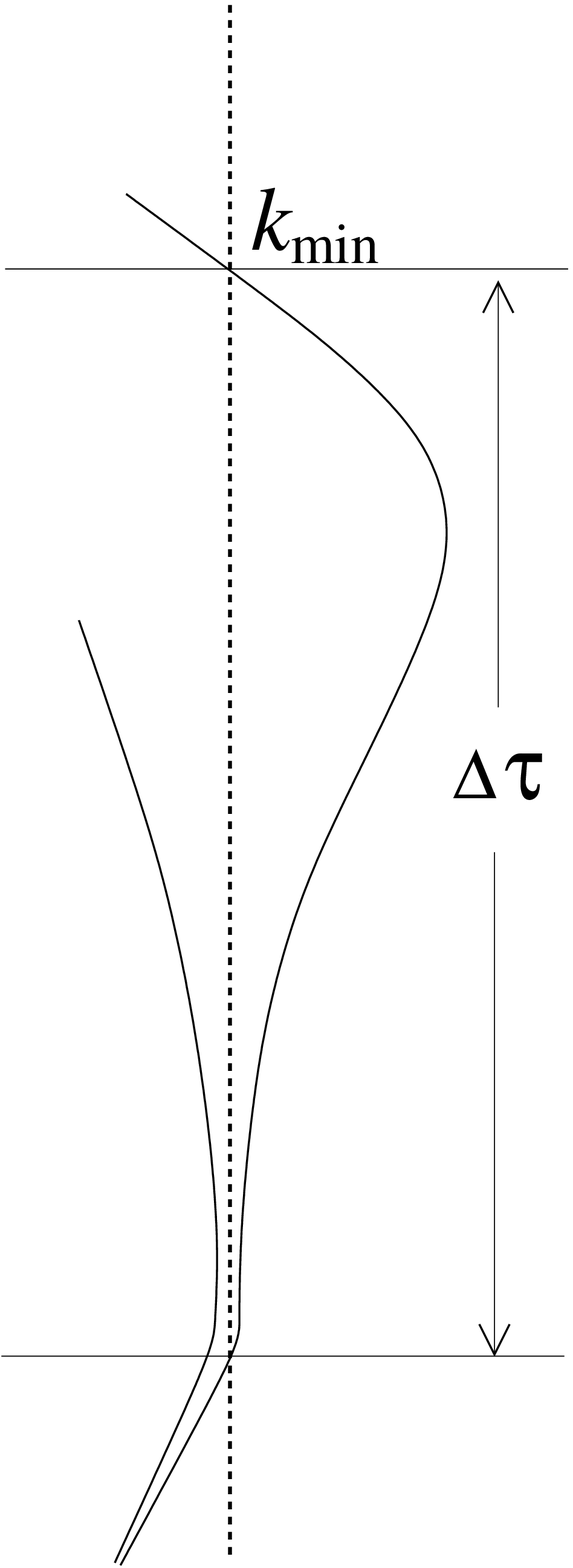}
   \hspace{1cm}
   \includegraphics[angle=0,width=3cm]{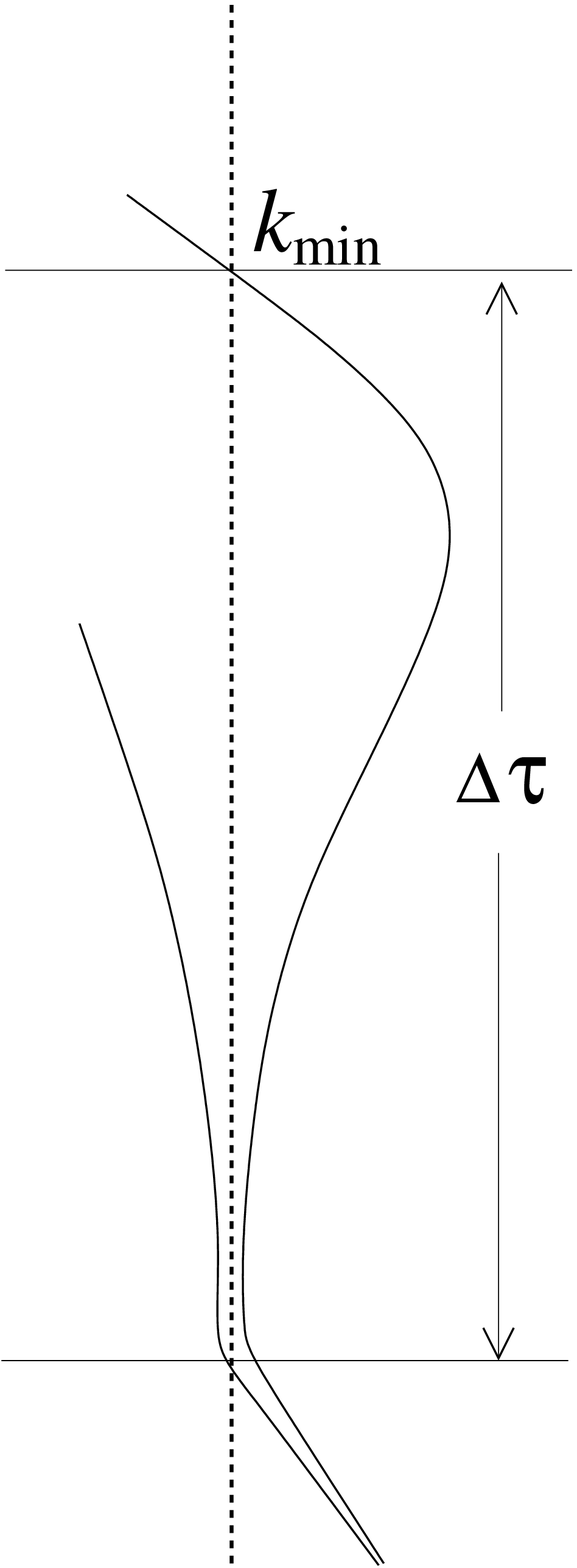}
\caption{\label{trajectories} Trajectories of the
particle and partner in the relativistic and dispersive
cases. The density of states is always proportional to
the
PG time
lapse
$\D \tau$. In the relativistic case (left) this is the time
the particle spends outside the horizon and below the
hard cutoff. In the super-luminal case (center) it is simply
the time the particle spends outside the horizon. In
the sub-luminal case (right) it is the time between the horizon
crossings of the partner and the particle.}
\end{figure}

The
relevant
mode density is the inverse of the frequency gap $\Delta\omega$
that separates two neighboring modes which are orthogonal with
respect to the one-particle Hilbert space inner product
restricted to
the black hole exterior.
As shown in the Appendix, in the WKB
approximation, this
mode density
is given by
\beq n(\omega,\ell)=(2\pi\kappa)^{-1}\ln(k_{\rm max}/k_{\rm min}),
\label{nlnk}\eeq
where $k_{\rm max}$ is the cutoff wavevector
$\L$ and $k_{\rm min}$ is the
value of $k_{\o\ell}(\rho)$  when the trajectory crosses the horizon after having reflected from the angular momentum barrier
(see Fig.~\ref{trajectories}).
In the near horizon region,
the wavevector decreases exponentially
as $\exp(-\k \tau)$
from $k_{\rm max}$ to $k_{\rm min}$ along the trajectory,
thereby allowing to write (\ref{nlnk}) as $n(\omega,\ell)=\Delta \tau /2\pi $.

We now explain why it suffices to restrict attention to the near
horizon region. At the turning points of the classical
trajectories the dispersion relation (\ref{reldisp}) implies
\beq 1-v_{\rm t.p.}^2  \simeq(\omega/p)^2, \label{vtp}\eeq
where $p$ is the transverse momentum defined by
\beq p^2=\ell(\ell+1)/r^2.\label{pt} \eeq
Since (as it turns out) the entropy is dominated by the high
angular momentum modes with $p\gg\kappa$, and the contribution of
high frequencies to the entropy is exponentially suppressed by
$\exp(-2\pi\omega/\kappa)$, we have $(1-v^2)_{\rm tp}\ll 1$.
It therefore suffices to use the linear
approximation (\ref{kappa}), and to set $r=r_H$ in the
definition (\ref{pt}) of the transverse momentum $p$.

 The dispersion relation
(\ref{reldisp}) at the horizon yields $k_{\rm min} = (p^2
-\omega^2)/2\omega$. The mode density (\ref{nlnk}) in the presence
of the cut-off is therefore given by
\beq n_{\rm cut}(\o,\ell)=(2\pi\k)^{-1}\ln\left(
\frac{2\o\L}{p^2-\o^2}\right)\label{ncut}\eeq
The condition that $k_{\rm min}$ be less than $k_{\rm max}=\L$
constrains $p$ to be less than
\beq p_{\rm max}=\sqrt{2\omega \L + \omega^2}, \label{pmax}\eeq
thereby eliminating the quadratic divergence
in (\ref{Satmo})
that comes from the
angular momentum sum. For transverse momenta $p$ close to $\o$ the
contribution to the entropy is not well determined by our
near horizon approximation since, as shown by (\ref{vtp}), the
turning point is not near to the horizon. However, these momenta
make a negligible contribution to the entropy. Thus we simply
adopt the lower limit $p_{\rm min}=\o$ that is imposed by the
condition $k_{\rm min}>0$, which is
obeyed by
any mode that
is initially outgoing.

Since the atmosphere entropy is dominated by large angular momenta
we can replace the sum over $\ell$ in (\ref{Satmo}) by an integral
over $p^2$,
\beq \sum_\ell (2\ell+1)\rightarrow r_H^2\int dp^2.
\label{ellsum}\eeq
Then inserting (\ref{ncut}) for the mode density into (\ref{Satmo})
yields
\beq S_{\rm atmo,\,  cut} = A\times (\kappa \L)\times
(2\pi)^{-4}
\int_0^\infty dx\,  x\,  s_{\rm
therm}(x),\label{Scut}\eeq
where $A$ is the horizon area $4\pi r_H^2$ and $x=2\pi\omega/\k$.
(The integral is equal $3\zeta(3)\simeq 3.6$.) This expression gives
the entropy of the black hole atmosphere when regulated by a hard
momentum cut-off imposed in the freely falling frame that is at rest
at infinity, evaluated using the WKB
approximation. We now comment on several aspects of this result.\\

\noindent{\it Frame dependence}---Had we imposed the cut-off $\L$
in a general free-fall frame the result would have been the same as
(\ref{Scut}), except with $\k$ replaced by $\k_v$ defined in
(\ref{kappa}) or equivalently (\ref{kvk}). Thus the entropy
actually depends on the free-fall frame in which the cut-off
is defined. This is to be expected, since the cut-off prescription
is not Lorentz invariant.\\

\noindent{\it Universal dependence on
transverse momentum cutoff}---Although the
entropy depends on the free-fall frame adopted for the
cut-off $\L$, it scales universally
with the maximum angular momentum as
\beq \ell_{\rm max}^2\sim A\, \bar{p}^2_{\rm max}, \eeq
with $\bar{p}_{\rm max}= \sqrt{2\k_v\L}$
the value of $p_{\rm max}$ (\ref{pmax}) evaluated at $\o=\k_v$
(where most of the entropy resides).
Thus each unit of area
$(\bar{p}_{\rm max})^{-2}$ carries of order one bit of entropy.
Put differently, the entropy scales as the number of
angular momentum states below the cutoff.
The same scaling is found in the case of a
dispersive cutoff in the next section.\\

\noindent{\it Near-horizon approximation}---As explained after
(\ref{pt}), the near-horizon approximation is justified for modes
with $p\gg\k$. We can now see that this is satisfied for the
dominant
modes, since $\bar{p}_{\rm max}\sim(\k\L)^{1/2}\gg\k$.\\

\noindent{\it Anisotropy of momentum cutoff}---Since $\bar{p}_{\rm
max}$ is much smaller than the cut-off $\L$, the entropy would be
unchanged were the cut-off $\L$ imposed isotropically
on the norm of the momentum
rather than on the radial component alone.\\

\noindent{\it Mode density for dominant angular momenta}---Since 
the entropy scales as the number of angular momentum
states, the number of modes contributing to
the entropy at fixed angular momentum $(\ell,m)$ is of order one
for the dominating values of $\ell$. Indeed, evaluating the mode
density (\ref{ncut}) for $p$ parametrized as a fraction of the
maximum value, $\bar{p}_{\rm max}/N$, yields $n_{\rm cut}=(\ln
N)/2\pi\k$. Since thermal suppression by the Boltmann factor limits
the effective frequency range to $\D\o\sim \k/2\pi$, the number of
contributing modes is $\sim n_{\rm cut}\D\o\sim (\ln N)/(2\pi)^2$.
Since most of the entropy comes from $N$ close to unity, only a
``fraction of a mode" contributes for each angular momentum.
In this regime $n_{\rm cut}\D\o$
should be interpreted as the {\it probability}
to find the corresponding particle in
the atmosphere outside the horizon.\\

\noindent{\it Comparison with ``brick wall" model}---The entropy
(\ref{Scut}) scales as $\k\L$, linearly with the momentum cutoff
$\L$.  This scaling differs from that of the entropy regulated
with a ``brick wall" cutoff~\cite{'tHooft:1984re}, i.e. a reflecting
boundary condition at a proper distance $\epsilon$ from the
horizon, which scales quadratically as $\epsilon^{-2}$. The
difference arises only because different frames have been used to
measure the cutoff scales: we measured $\L$ on a time slice
orthogonal to the free-fall worldlines, whereas $\epsilon$ was
measured in Ref.~\cite{'tHooft:1984re} along a static time slice,
orthogonal to the Killing vector. On the free-fall time slice, the
same brick wall lies at a much smaller proper distance
$\d=\kappa\epsilon^2/2$ from the horizon. Hence when expressed in
terms of $\d$, the brick wall entropy scales as $\d^{-1}$, in
agreement with what we found.

The agreement in overall scaling with the cutoff is a feature of
the total entropy, after integrating over all frequencies and
transverse momenta. This agreement arises because the dominant
contributions scale the same way, despite the fact that the
physical nature of the two cutoffs is different. In particular,
the brick wall cuts off all modes at the same {\it location},
whereas we cut off all modes at the same {\it momentum}. In our
case the cutoff location is not sharp, but can be characterized
using the WKB approximation. Its value then depends on the
frequency and angular momentum. The angular momentum dependence is
negligible because, as noted after equation (\ref{pmax}), $p\ll\L$ for
the frequencies $\o\sim\k$ that dominate the total entropy. The
frequency dependence on the other hand is significant. Using the
dispersion relation (\ref{reldisp}) one finds near the horizon
$k\simeq\o/\k(\rho-\rho_H)$, so that the WKB position cutoff is
$\o/\k\L$. However, this $\o$ dependence does not affect the total
entropy scaling since frequencies much greater than $\k$ are
thermally suppressed. A further negligible difference with the
brick wall evaluation arises because we cut off the WKB
trajectories at the final horizon crossing, whereas the brick wall
trajectories begin and end at the same radius, as explained in
more detail in the Appendix.

\section{Dispersion induced cut-off}
We now remove the high wavevector cutoff $\L$ and instead
introduce dispersion into the field propagation by including
higher space derivative terms suppressed by powers of $\L $. Both
super-luminal and sub-luminal dispersion will be considered.
As explained above, it did not matter much whether the hard
cut-off was isotropic or only in the radial direction. In the
dispersive case the same will be true, at least in the
super-luminal case. Whether it is true in the sub-luminal case
will depend upon an open question to be discussed below.

The
relativistic dispersion relation between frequency
$\Omega$ and
wavenumber $q$ for a massless field in flat spacetime is $\Omega^2
= q^2$, where $q^2={\bf q}\cdot{\bf q}$.
We consider an
isotropic
UV-modified dispersion relation of the form
\beq \Omega^2 = F^2(q)= \left(q \pm \textstyle{\frac{1}{n}}q^n/\L ^{n-1}\right)^2,
\label{F}\eeq
chosen for convenience so that $F^2(q)$ is a perfect square.
The quantity $\L $ sets the scale for the nonlinear dispersion,
and we assume $n>1$.
At low wavenumbers $q\ll \L $ the dispersion is relativistic. At
high wavenumbers the group velocity
$v_g = \partial_q\Omega$ is super-luminal for the plus sign and
sub-luminal for the minus sign. At $q=\L $ we have $\Omega(\L
)=(1\pm \textstyle{\frac{1}{n}})\L $ and $v_g(\L )=1\pm 1$. We now adopt units with
$\L =1$.

The dispersion relation has the given form only in a particular
rest frame. As in the case of the hard cutoff we choose this to be
the free-fall frame that is at rest at infinity.
As in the nondispersive case,
we can decompose the scalar field in the form (\ref{partialwaves})
and use the WKB approximation for the radial functions
$u_{\o\ell}$. The positive free fall frequency outgoing ($k>0$)
mode with angular momentum $\ell$ satisfies the dispersion
relation
\bea
\Omega =
 \omega -v(\rho)k
 &=& (k^2 + p^2)^{\textstyle{\frac{1}{2}}} \pm \textstyle{\frac{1}{n}}
(k^2 + p^2)^{\textstyle{\frac{n}{2}}}\\
&\approx& k + p^2/2k \pm k^n/n, \label{disprel} \eea
where the approximate form in the last line holds when
$p\ll k$
and $k \ll \L  = 1$.
As can be verified a posteriori, this approximation is valid in
the present context,
as in the hard cutoff case. Moreover,
the $r$ dependence in $p^2=\ell(\ell+1)/r^2$ can be safely
replaced by the constant value $r_H$.

As before we take the field to be in the free fall vacuum state. A num-
ber of studies have demonstrated that in this state, to leading order in
 $\kappa/\L$, models with UV dispersion reproduce the relativistic result that the
outgoing modes at infinity are thermally populated at the Hawking
temperature~\cite{Unruh:1994je,Brout:1995wp,Corley:1996ar,Jacobson:1999zk}. To evaluate the entanglement entropy we need to
know whether thermality extends to the near-horizon modes that contribute
to the entropy. In the Appendix we deduce the quantum field theoretic
description of the near-horizon free fall vacuum, generalizing to four
dimensions the treatment of Ref.~\cite{Balbinot:2006ua}.
The state is a product of two-mode squeezed
states that are thermal when reduced to one mode.
We also discuss in the Appendix
how the WKB approximation can be used to identify those entangled modes localized on either side of the horizon. In this section we shall assume that this WKB treatment provides an accurate evaluation of their density.

Under the above assumptions, the atmosphere entropy takes the same
form (\ref{Satmo}) as before, but with a different density of
modes $n_{\rm disp}(\omega,\ell)$.
As shown in the Appendix, the mode density is again
given by (\ref{nlnk}).
As in the relativistic case, the wavevector $k_{\rm min}$
corresponds to the point where the particle trajectory finally
crosses the horizon.
In the WKB approximation,
the {\it only} place where the dispersion
enters the calculation of the atmosphere entropy is in the
determination the initial wavevector $k_{\rm max}$.

In the super-luminal case the outgoing positive frequency modes
emerge from behind the horizon after parting ways with their
negative frequency partners which always remain behind the horizon
(see Fig.~\ref{trajectories}). Therefore the relevant trajectory for
determining the density of states to be included in the entropy
extends between the two horizon crossing points of the positive
frequency trajectory, and $k_{\rm max}$ is the value of $k$ at the
first (outward) horizon crossing. The mode density at fixed
$\omega$ and $\ell$ is thus finite, and the sum over
$\ell$ is cut off when
the turning point drops below the horizon and
the two horizon crossing points merge.
The atmosphere entropy is thus finite. We now discuss the
determination of $k_{\rm max}$ in the sub-luminal case, before
computing the entropy for both cases together.

In the sub-luminal case, the positive and negative frequency
outgoing modes are initially infalling with respect to the static
frame. For high enough frequency or low enough angular momentum
the positive frequency modes convert to outward propagation at a
turning point just outside the horizon (see Fig.~\ref{trajectories}),
followed by a second turning point at the angular momentum barrier
where they begin to fall back in. For lower frequency or higher
angular momentum the pair of turning points disappears and the
particle just follows the partner straight across the horizon with
a small time lag. Since
the entanglement of these
modes is the source of the entropy, we shall identify $k_{\rm
max}$ the value of $k$ for the positive frequency particle at the
time the negative frequency  partner
crosses the horizon, leaving
the particle stranded outside. This notion requires us to
determine what is the particle momentum at a given PG time when
the partner has a given momentum $k$. This determination is
impossible if we refer only to Hamilton's equations, but
in quantum mechanics the
positive and negative frequency modes are linked through the
specification of the in-modes which define the
vacuum state.
As shown in the appendix,
by building wave packets of in-modes one can see
that the particle and partner possess the {\it same} mean
momentum at the same PG time. We therefore identify $k_{\rm max}$
as the value of the momentum when the {\it partner} crosses the
horizon,
see Fig.~\ref{trajectories}.

The mode density $n(\o,\ell)$ of (\ref{nlnk})  is thus again finite,
since $k_{\rm max}/k_{\rm min}$ is finite. There remains a
potential divergence in the sum over angular momenta however
since, unlike in the super-luminal case, this ratio remains
strictly greater than unity. This is because the partner always
crosses the horizon a finite time before the particle, as
explained in the Appendix.  The mode density therefore remains
non-zero no matter how large $\ell$ becomes, and it may not
approach zero fast enough for the sum over $\ell$ to converge. We
will examine this potential divergence more closely later on.

The
values of $k_{\rm max}$ and $k_{\rm min}$ are given by
appropriate roots of the dispersion relation (\ref{disprel})
evaluated at the horizon,
\beq
\omega = p^2/2k \pm k^n/n.
\eeq
The upper and lower signs are for the super- and sub-luminal cases
respectively. This can be expressed in terms of the scaled
variable $\tilde{k}=k/|\omega|^{\frac{1}{n}}$ as
\beq \s\tilde{k} \mp \textstyle{\frac{1}{n}}\tilde{k}^{n+1} =p^2/2|\omega|^{1+\frac{1}{n}},
\label{tqhorizon}
\eeq
where $\s$ is the sign of $\o$, positive for particle and negative
for partner. The ratio $k_{\rm max}/k_{\rm min}$ is the same as
the ratio of the $\tilde{k}$'s, hence it depends upon $\omega$ and
$p$ only through the combination $p^2/\omega^{1+\frac{1}{n}}$.

Following
the above discussion, the mode density $n(\omega,\ell)$ in the super-luminal case is determined by the log of the ratio of
the two roots $\tilde{k}_{\rm max}$ and $\tilde{k}_{\rm min}$ of
(\ref{tqhorizon}) for $\s=+1$ (particles)
and the upper (negative) sign.
The maximum transverse momentum
is determined
by the maximum of the left hand side of (\ref{tqhorizon}).
It is given by
\beq
p_{\rm max}=\sqrt{2\o\L}\left(\frac{\o}{\L}\right)^{\frac{1}{2n}}
\left(\frac{n}{n+1}\right)^{\frac{n+1}{2n}},
\label{pmaxdisp}\eeq
where the explicit dependence on the momentum scale $\L$
has been restored.
This is essentially the same as in the case of the hard cutoff
(\ref{pmax}) in the limit $n\rightarrow\infty$, when the dispersion
turns on suddenly at $k=\L$.
When $n$ is finite, $p_{\rm max}$ scales with a lower power
of the ``new physics" scale $\Lambda$ than for the hard cutoff.
Therefore the  atmosphere entropy will be similarly reduced.

\begin{figure}
 \includegraphics[angle=0,width=10cm]{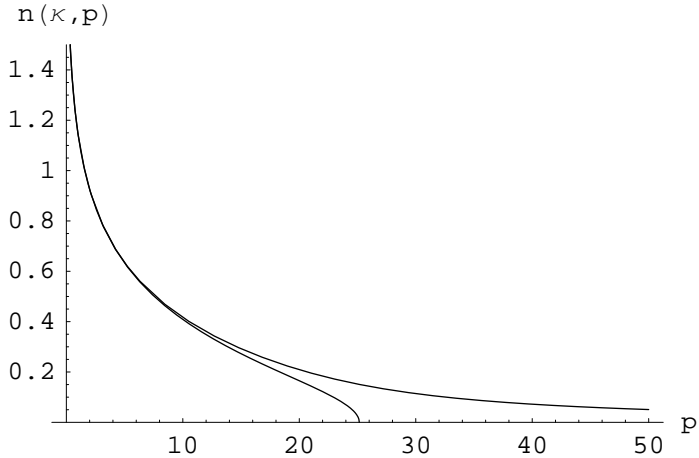}\\
\caption{\label{modedensity} Plot of the mode density (\ref{nL}) vs.
the transverse momentum $p$ at fixed frequency $\o=\kappa$, for
sub-luminal (upper curve) and super-luminal (lower curve)
dispersion, with $\Lambda/\kappa=10^4$ and $n=3$, in units with
$\kappa=1$.}
\end{figure}

In the sub-luminal case the lower sign applies, and $\tilde{k}_{\rm
min}$ is the root for $\s=+1$ (particle)
while $\tilde{k}_{\rm max}$ is the
root for $\s=-1$ (partner). In both cases, the mode density
can be expressed as
\beq n_{\rm disp}(\omega,\ell)
=\frac{1}{2\pi\kappa} \ln(\tilde{k}_{\rm
max}/\tilde{k}_{\rm min})=\frac{1}{2\pi\kappa}\, {\cal
N}_n^\pm\left(\frac{p^2}{\o\L(\o/\L)^{\frac{1}{n}}}\right),\label{nL}\eeq
where ${\cal N}_n^\pm$ is a function of
its dimensionless argument that cannot be
written in closed form.
In Fig.~\ref{modedensity} we plot this function
for both sub-luminal (${\cal N}_n^-$)
and super-luminal (${\cal N}_n^+$) dispersion
in the case $n=3$. Notice that while the
cutoff scale is taken to be $\L=10^4\k$,
the
maximal value of the
transverse momentum in the super-luminal case
is only $\simeq 25\k$.

Using the mode density just obtained, the entropy (\ref{Satmo})
(again using (\ref{ellsum})) evaluates to
\beq S_{\rm atmo,\,  disp} = A\times \kappa \L  \times
\left({ \kappa\over \L }\right)^{\frac{1}{n}}
\times C^\pm_n,\label{Sdisp}\eeq
where the dependence on $\L$ has been restored. The factor
$C^\pm_n$ is a constant defined by
\beq C^\pm_n=
 (2\pi)^{-4 - \frac{1}{n}}
\Bigl(\int_0^{y_{{\rm max},n,\pm}} dy \,
{\cal N}_n^\pm(y)\Bigr) \Bigl(\int_0^\infty dx\,  x^{1+\frac{1}{n}}\,
s_{\rm therm}(x)\Bigr), \label{Cn} \eeq
where $y$ is the dimensionless variable
$p^2/(2\omega\L(\o/\L)^{\frac{1}{n}})$. The upper limit of integration
$y_{{\rm max},n,+}$ in the super-luminal case  is
the maximum
value of the left hand side of (\ref{tqhorizon}), yielding a
finite total entropy within the WKB treatment we have adopted. In
the sub-luminal case $y_{{\rm max},n,-}=\infty$, as explained
above, hence the result depends on the asymptotic mode density at
momenta larger than $\L$.

The expression (\ref{Sdisp}) gives the atmosphere entropy in the
dispersive case.
As in the case of the hard cutoff, the entropy scales
universally with the square of the angular momentum cutoff,
i.e.\ as $\ell_{\rm max}^2\sim A\, \bar{p}^2_{\rm max}$.
In the present case, $\bar{p}_{\rm max}$ is the
value of (\ref{pmaxdisp}) at $\o=\k$
(where most of the entropy resides).
Comparing with our previous result (\ref{Scut})
for the case of a hard cutoff, we note that the
scale $\L$ enters with
a different power which depends on $n$. In the limit
$n\rightarrow\infty$ one recovers the hard cutoff result, at least
in the super-luminal case. This happens because for large $n$ the
dispersion turns on abruptly at $k=\L$.
\\

\noindent{\it Asymptotic mode density in sub-luminal case}---The
absence of an upper limit to the angular momentum sum in the
sub-luminal case means that the result depends on the behavior of
the dispersion relation at momenta higher than $\L$. The
dispersion relation (\ref{F}) was introduced to parametrize only
the first order correction to the relativistic dispersion. To
analyze the mode density in the regime of very high momenta the
first term in the Taylor expansion is inadequate, so we allow for
a general dispersion relation of the form $(\o-vk)^2=F^2$ where
$F(k,p)$ is now an arbitrary function giving rise to sub-luminal
propagation.

At fixed $k$ and $p$, as shown in the Appendix,
 the dispersion relation implies that the
particle and partner locations near the horizon are separated by
$\Delta\rho=2\omega/(\kappa k)$, with the particle at larger
$\rho$.
The mode density is determined by the time
between the horizon crossings of the partner and the particle,
$\Delta \tau \approx|\Delta\rho/(d\rho/d\tau)|= 2\omega/(\kappa
k(1-F_{,k}))$. It is hence given approximately by
\beq n_{\rm sub}(\o,\ell)=  \frac{\o}{\pi\kappa k(1-F_{,k})}, \eeq
where $k$ satisfies the dispersion relation at the horizon,
$\omega+k=F(k,p)$. As long as $F\gg\o$, we have $k\simeq F(k,p)$
hence $n_{\rm sub}$ behaves as $\sim\o/F$ since for the
sub-luminal case $F_{,k}$ is assumed less than one. The number of
modes in the relevant thermal frequency range $\D\o\sim\k$ is
therefore $\sim \k/F\ll1$, assuming $F\gg\k$. That is, there is
only a tiny probability to find a particle with large transverse
momentum. The WKB approximation is
probably
unreliable in this regime, so
our methods are unable to evaluate the contribution of these modes
to the entropy.

\section{Discussion}

We have found that a freely falling hard momentum cutoff produces an
entanglement entropy that scales as $A\, \k_v\L$ with the cutoff
$\L$ and the surface gravity $\k_v$ defined in a given free fall
frame. When the frame is taken to be at rest at infinity $\k_v$
coincides with the usual surface gravity $\k$. With no cutoff but
instead sub- or super-luminal dispersion with index $n$ the scaling
has an additional factor of $(\k/\L)^{1/n}$.  If the freely falling
cutoff $\L$ is of the order of the Planck mass $M_p$ this entropy is
negligible when compared to the Bekenstein-Hawking entropy $S_{\rm
BH}\sim AM_p^2$ (for macroscopic black holes). This would
conveniently eliminate the species problem, i.e. the non-universal
contribution of the entanglement entropy to the black hole entropy,
without the need to suppose a  relation between the entanglement
entropy and renormalization of the low energy effective Newton
constant.

If instead the cutoff were given by $\L=M_p^2/\k$ in the freely
falling frame the scaling of the entanglement entropy would agree
with $S_{\rm BH}$. This is equivalent to saying that the cutoff is
actually $M_p$ in the static frame. While this might be considered
natural from
a ``geometric" viewpoint, it seems unnatural from the viewpoint of
the equivalence principle, which suggests that the cutoff should
be a property of the freely falling vacuum. However, the
equivalence principle would not forbid a cutoff arising from
quantum gravity effects sensitive to the global properties of the
background. In fact, there is some evidence for this.

Strong
gravitational interactions between ingoing and outgoing
quanta
arise when the
invariant mass squared
reaches $M_p^2$~\cite{'tHooft:1984re}.
The radiative corrections associated with these interactions
in the free fall vacuum outside the black hole
were studied in Ref.~\cite{Parentani:2000ts,Parentani:2002mb}.
Beyond the local effects that would also be present in
the vacuum of flat
spacetime, it was found that there are further contributions,
dominated by in-modes with
Killing frequency $\omega_{\rm in}\sim\k$.
These corrections
induce a
dissipating cutoff on the
(backward) propagation of outgoing modes
at the location where
their frequency is $M_p^2/\k$ in the
freely falling frame,
i.e.\ $M_p$ in the static frame.
While the significance of that preliminary
analysis remains unclear, it does indicate a way in which the
cutoff entanglement entropy might account for the
Bekenstein-Hawking entropy.

It would be interesting to extend the considerations of the
present paper to the cases of de Sitter and Rindler horizons.
In the near-horizon limit these are identical to a black
hole horizon, and a strong case can be made that
they too posses a universal entropy density
$1/4L_p^2$~\cite{Jacobson:2003wv}.
On the other hand, they differ in the nature of their
``observer-dependence" and,
perhaps more importantly, in the nature of the
globally stationary free-fall frames that they admit.
We leave the study of the regularized entanglement
entropy for such horizons to future work.

We would like to reiterate that the WKB
approach employed in this paper cannot provide conclusive results
regarding the finiteness or scaling of the entanglement entropy
in the dispersive cases, and it is particularly inconclusive in the
sub-luminal case.
The uncertainty arises since the mode density
for the relevant modes is smaller than unity, so that
we are not in the regime where a WKB treatment is
guaranteed to be reliable.
Therefore only a field theory analysis of the reduced
density matrix outside the horizon would permit definitive
conclusions to be reached.

In this paper we have studied
the entanglement entropy
between the degrees of freedom
on either side of
the sharp, low frequency
horizon on a fixed black hole background.
In the relativistic
case, this horizon is a strict causal barrier,
but in the dispersive cases it is
dynamically distinguished only  in the low frequency
limit.
(For sub-luminal dispersion the
location of the barrier is ``frequency dependent",
while in the super-luminal case there is no
barrier at all.)
While this entanglement entropy is  interesting,
a more important, physical
problem is to evaluate the
 ``actual" thermodynamic or statistical entropy.
The physically relevant definition of entanglement
is not so clear in the dispersive cases,
 and in fact is unclear
even in the
relativistic case,
once quantum fluctuations
of the horizon are admitted.
Presumably,
the definition should
ultimately be provided by a specified set of
physical observers, interacting with the system
in a definite manner.\\

\noindent {\it Note added}:
While this paper was in press we became aware of a paper
by Frolov and Novikov~\cite{Frolov:1993ym} in which a regulated black hole
entanglement entropy for a massless scalar field was computed.
A method similar to the one used here using WKB trajectories
was employed to identify the modes contributing to the entropy.
The cutoff was imposed by including only contributions
from modes outside a certain minimum distance from the
horizon, motivated by computations of quantum horizon fluctuations.
The minimum distance was taken to be of the order the Planck length
as measured on a static slice, so that the entropy scales
with the area in Planck units.

\section*{Acknowledgments}

This work was supported in part by the National Science Foundation under grants PHY-0300710 and PHY-0601800 through the University of Maryland,  by the CNRS at the Institut d'Astrophysique de Paris,
and by NATO grant CLG-979723. We are
also grateful for the hospitality of both
the University of Tours
and
Perimeter Institute,  where some of this
work was done.

\appendix
\renewcommand{\theequation}{A.\arabic{equation}}
  \setcounter{equation}{0}  
\section*{Appendix}

In this appendix we spell out  the field theoretic
rationale for our WKB evaluation of the
mode density relevant to the entanglement entropy.
The central ingredient is a description of the
correlated structure of the near-horizon free-fall
vacuum in terms of
modes written in momentum space.
These results could be useful in a future approach
to computing the entanglement entropy without recourse
to the WKB approximation.

\subsection*{Action \& field equation}

We consider a massless scalar field in the spacetime with line
element (\ref{ds2}).
It is governed by the action
\bea
I&=& \frac{1}{2}\int d^4x\,  \sqrt{-g}\, g^{\mu\nu}\, \partial_\mu\varphi\, \partial_\nu\varphi \nonumber
\\
&=&\frac{1}{2}\int d\tau\, d\rho\, d\Omega\;
\Bigl\{r^2\Bigl[\bigl((\partial_t+v\partial_\rho)\varphi\bigr)^2 -
(\partial_\rho\varphi)^2\Bigr]
\nonumber\\
&&~~~~~~~~~~~~~~~~~~~~~~~~~~~~
-(\partial_\theta\varphi)^2-\frac{1}{\sin^2\theta}(\partial_\phi\varphi)^2\Bigr\}.
\label{action}
\eea
The field equation arising from this action is
\beq
(\partial_\tau+\partial_\rho v)r^2(\partial_\tau
+v\partial_\rho)\varphi-\partial_\rho r^2\partial_\rho\varphi
-\frac{1}{\sin\theta}\partial_\theta \sin\theta\,  \partial_\theta\varphi-\frac{1}{\sin^2\theta}\partial_\phi^2\varphi=0.
\eeq

\subsection*{Modes}
Modes of fixed frequency and angular
momentum take the form
\beq \varphi_{\omega\ell m} = \exp(-i\omega \tau)r^{-1}
u_{\omega\ell}(\rho) Y_{\ell m}(\theta,\phi).
\label{partialwaves2}\eeq
To find the equation satisfied by the mode function $u_{\omega\ell}$
we compute
\beq
\partial_\rho r^2\partial_\rho r^{-1}=r\partial_\rho^2-r'' ,
\eeq
and
\beq
(-i\o+\partial_\rho v)r^2(-i\o+v\partial_\rho)r^{-1}=r(-i\o+\partial_\rho v)(-i\o+v\partial_\rho)-(v^2r')' ,
\eeq
where the prime $'$ denotes $\partial_\rho$.
The mode equation is thus
\beq
\Bigl[(-i\o+\partial_\rho v)(-i\o+v\partial_\rho)-\partial_\rho^2+V_\ell(\rho)\Bigr]u_{\o \ell}=0 ,
\label{modeq}\eeq
with
\beq
V_\ell(\rho)= \frac{(1-v^2)r''}{r}-\frac{2vv'r'}{r}+\frac{\ell(\ell+1)}{r^2} .
\label{Vell}\eeq
This is written for the general line element (\ref{ds2}).
With the
Painlev\'e-Gullstrand (PG) coordinates
for the Schwarzschild metric
we would have $r(\rho)=\rho$,
so the first term in $V_\ell$ vanishes and the second term becomes just
$-2vv'/r$.

\subsection*{Near-horizon approximation}
We are interested in modes propagating very near the horizon since these
are the ones that contribute most of the entanglement entropy.
Expanding about the horizon in $x=\rho-\rho_H$ we have
\bea
v(\rho)&=&-1+\k_v x +\cdots\label{vexp}\\
V_\ell(\rho)&=&\frac{2\k_vr'(0)}{r_H}+\frac{\ell(\ell+1)}{r_H^2}+V'(0) x +\cdots
\eea
where $\k_v=v'(0)$ is the
surface gravity. The higher order terms in
$x$ are negligible in the near-horizon region, and the $O(x)$ term is only significant
when multiplied by a large coefficient. The derivatives $\partial_\rho$
in the first term of (\ref{modeq}) provide large
coefficients for the modes of interest, due to the large redshift at the horizon,
so we retain the $O(x)$ contribution to the velocity (\ref{vexp}) in that term.
Since the entanglement entropy is dominated by large angular momenta,
we can drop from $V_\ell$ all but the $\ell$-dependent term. In the near-horizon
region we can set this term equal to its value at the horizon,
$V_\ell =p^2$, where we have introduced
the notation
\beq p^2=\ell(\ell+1)/r_H^2 \label{p}\eeq
for the transverse momentum squared evaluated at the horizon.

\subsection*{Momentum space}

The momentum space modes $\tu(k)$ are related to $u(\rho)$ by
Fourier transform,
\beq
u(\rho)=(2\pi)^{-1/2}\int dk\, \tu(k)\exp(ik\rho).
\eeq
In momentum space the mode equation (\ref{modeq}) becomes
\beq
(\o-k \hv)(\o-\hv k) \tu_{\o\ell}= \bigl(k^2+\hat{V}_\ell\bigr)\tu_{\o \ell},
\label{modeqk}\eeq
where $\hv=v(\hat{\rho})$ and $\hat{V}_\ell= V_l(\hat{\rho})$ with
$\hat{\rho}=i\partial_k$. In the near-horizon approximation, and for large $\ell$
this mode equation takes the simpler form
\beq  (\o-k \hv)(\o-\hv k) \tu_{\o\ell}=F^2_\ell(k) \tu_{\o \ell},
\label{modeqknear}\eeq
where now
\beq
\hv =-1+\k_v\hat{\rho}= -1+i\k_v\partial_\rho,
\eeq
and
\beq
F^2_\ell(k)= k^2+p^2.
\label{Frel}\eeq
To simplify the notation we now drop the subscript $v$ on $\k$.

Thanks to the identities
\bea
(\o-\hv k)k^{-i\o/\k-1}e^{-ik/\k} \, u
&=&  k^{-i\o/\k-1}e^{-ik/\k} \, (-i\k k \partial_k u),
\label{useful}\\
(\o-k\hv)(\o-\hv k)k^{-i\o/\k-1}e^{-ik/\k}\,  u
&=& k^{-i\o/\k-1}e^{-ik/\k}\, (-\k^2k^2\partial_k^2 u), \quad\quad
\eea
the  $\o$-dependence of the solutions
to (\ref{modeqknear}) can be factored out,
so the solutions take the form
\beq
 \tu_{\o\ell}=k^{-i\o/\k-1}e^{-ik/\k}\, \chi_\ell(k),\label{factored}
\eeq where $\chi_{\ell}$ obeys \bea
-\k^2 k^2\partial_k^2\chi_\ell &=& F^2_\ell\chi_\ell, \label{chi}\\
&=& (k^2+p^2)\chi_\ell.\nonumber
\eea
For $k\gg p$ the solutions are $\exp(\pm ik/\k)$, while for $k\ll p$
they are $k^\a$ with $\a=\half\pm\sqrt{\quarter -p^2/\k^2}$. Only
the first case is relevant for the modes outside the horizon
however, since $k_{\rm min}\simeq p^2/\o$ which is much greater than
$p$ for the modes relevant to the entropy.

It so happens that (\ref{chi})
is closely related to Bessel's equation, and is exactly solvable in
terms of Bessel functions:
\beq \chi_\ell(k)= \sqrt{k}\, {\cal C}_\n(k), \qquad
\n=\sqrt{\quarter-p^2/\k^2}\label{Bessel}\eeq
where ${\cal C}=J,Y,H^{(1,2)}$.
As shown below these solutions
can be classified
according to whether $\chi_\ell$ has
positive or negative Wronskian
\beq
W= -i(\chi_\ell^*\partial_k\chi_\ell
-(\partial_k\chi_\ell^*)\chi_\ell). \eeq

\subsection*{Dispersion}

Now we consider modifying the field equation by the addition of
higher spatial derivative terms in the local rest frame defined by
the unit timelike vector field
$w=\partial_\tau +v(\rho)\partial_\rho$.
For the metric (\ref{ds2}) this frame is orthogonal to the
surfaces of constant $\tau$.

The dispersion of interest becomes important at high wavevectors,
where ``high" is specified by a momentum scale $\L \gg \kappa$.
 Since the modes of interest are only important near the horizon, it is
sufficient to study dispersion by a modification of the near-horizon
mode equation.  To be able to
handle arbitrary dispersion, it is
convenient to work in momentum
space~\cite{Brout:1995wp,Balbinot:2006ua}.  Allowing for
distinct modifications of
 (\ref{modeqknear})
 in the radial and transverse directions
yields the modified near-horizon mode equation
\beq (\o-k \hv)(\o-\hv k) \tu_{\o p}=F^2(k,p;\L) \tu_{\o p},
\label{modeqknearmod}\eeq
where
$F(k,p;\L)$ is a function determining
the dispersion, which is assumed to be relativistic
if $k, p\ll \L$.

As before,
the $\o$-dependence of the solutions to
(\ref{modeqknearmod}) can be totally factored out,
so the solutions again take the form (\ref{factored}),
\beq
 \tu_{\o p}=k^{-i\o/\k-1}e^{-ik/\k}\chi_{p,\L}(k),\label{factoreddisp}
\eeq
but
with $\chi_\ell$ replaced by $\chi_{p,\L}$.
The latter satisfies
(\ref{chi}) with the new $F^2$ function:
\beq -\k^2 k^2\partial_k^2\chi_{p,\L}=F^2(k; p,\L)\chi_{p,\L}.\label{chiL} \eeq

\subsection*{Klein-Gordon inner product}

On a constant $\tau$ slice the KG inner product in both the
relativistic case and in the presence of dispersion
is given by
\beq
\la \varphi_1,\varphi_2\ra=i\
\int d\Omega\, d\rho\,  r^2\,
\Bigl( \varphi_1^*(\partial_\tau+v\partial_\rho)\varphi_2-
((\partial_\tau+v\partial_\rho)\varphi_1^*)\varphi_2\Bigr),
\eeq
because the dispersion we consider only affects the spatial part of the action.
It is diagonal on angular momentum modes,
and for partial waves (\ref{partialwaves})
with the same angular momentum
it takes the form
\beq
\la \varphi_{\o'},\varphi_{\o}\ra=\int d\rho\,
\Bigl( u_{\o'}^*(\o+iv\partial_\rho)u_{\o}+
((\o'-iv\partial_\rho)u_{\o'})^*u_{\o}\Bigr).
\eeq
In  momentum space
it becomes
\beq
\la \varphi_{\o'},\varphi_{\o}\ra=\int dk \Bigl( \tu_{\o'}^*(\o-\hv k)\tu_\o+((\o'-\hv
k)\tu_{\o'})^*\tu_\o\Bigr). \label{KGk}
\eeq
For solutions factorized as in
(\ref{factored}) or in (\ref{factoreddisp}),
we make use of (\ref{useful}) to obtain
\beq \la \varphi_{\o'},\varphi_{\o}\ra=-i\k\int dk\,  k^{i\D\o/\k-1}
\bigl(\chi^*\partial_k\chi -(\partial_k\chi^*)\chi\bigr)
\label{KGDo}\eeq
where $\D\o=\o'-\o$.
In addition, for any solutions of (\ref{chi}) or its
dispersive modification (\ref{chiL}), the Wronskian of $\chi$ is a constant. Thus, {\it irrespective of angular momentum, dispersion,
or WKB approximation}, the KG overlap of two
modes in
the near-horizon region is proportional to
\beq \int dk
k^{i\D\o/\k-1} =\int d(\ln k) e^{i(\ln k) \D\o/\k}.
\label{overlap}
\eeq

\subsection*{Mode density
and classical trajectories}
To compute the entanglement entropy we need to know the density
$n(\o,\ell)$ of modes exterior to the horizon. This is just the
reciprocal of the gap in Killing frequency $\o$ that separates one
exterior mode from the nearest orthogonal exterior mode with the
same angular momentum. Orthogonality here is defined  with respect
to the one-particle Hilbert space inner product, which is defined by
the Klein-Gordon inner product. The mode density is thus the inverse
of the minimal $\Delta\omega$ for which the overlap (\ref{overlap})
vanishes, hence
\beq n(\omega,\ell)=(2\pi\kappa)^{-1}\ln(k_{\rm max}/k_{\rm min}).
\label{nlnkApp}\eeq
We implement the restriction to exterior modes
through the limits of integration $k_{\rm max}$ and $k_{\rm min}$, which are determined using the WKB approximation.

In both the relativistic and
dispersive cases $k_{\rm min}$ is the value of the WKB momentum
$k_{\o\ell}(\rho_H)$ at the horizon crossing where the
trajectory of
the
particle
falls back into the black hole after having
reflected from the angular momentum barrier (see
Fig.~\ref{trajectories}). The value of $k_{\rm max}$ is just the hard
cutoff $\L$ in the relativistic case, and in the dispersive case it is
determined by the passage of the particle (or its partner) across the horizon, as
explained in the text.

The dispersion relation
associated with the mode equation (\ref{modeqknearmod})
is
\beq
\o - v(\r)k = F(k, p; \L),
\label{nhdisp}
\eeq
where we took the positive root to describe outgoing modes.
The classical
trajectories satisfy Hamilton's equations,
\bea d\rho/d\tau&=&\partial\omega/\partial k ,
\label{Hamrho}\\
dk/d\tau &=&-\partial\omega/\partial\rho, \label{Hamk}\eea
where the Hamiltonian is given by
$\omega(\rho,k; \ell)$, solution of (\ref{nhdisp}).
In the near-horizon region, where $v(\rho)\approx -1+\k x$,
(\ref{Hamk}) implies the simple relation
\beq dk/d\tau = -\kappa k, \label{kdot}\eeq
both in the relativistic and dispersive cases. The wavevector $k$
therefore decreases exponentially as $\exp(-\k \tau)$ along the
trajectory, both before and after reflection from the angular momentum barrier
(which for large $\ell$ lies in the near-horizon region).
Notice that the same exponential law with $k > 0$ applies
to
the trajectories of the (outgoing but trapped)
partners characterized by the opposite Killing frequency $-\o$.
The sign of $k$ should therefore be thought as that of the FF frequency.

Using (\ref{kdot}), the mode density (\ref{nlnkApp}) can thus be reexpressed
in the simple form
\beq n(\omega,\ell)=(2\pi)^{-1}\Delta \tau(\omega,\ell), \label{nDtApp}\eeq
where $\Delta \tau(\omega,\ell)$ is the lapse of
 PG time
evaluated along
the relevant part of the classical trajectory. Two
near-horizon modes are
therefore orthogonal with respect to the KG product if
their frequencies are
separated by $\D\o= 2 \pi / \Delta \tau$, irrespectively of the
value of $p$ and the dispersion law $F$.\footnote{The linear growth in time
of the relevant mode density in (\ref{nDtApp}) is familiar from
the behavior of $\Sigma_{modes} = (\Delta t /2 \pi) \int d\o \rho(\o)$,
the number of modes involved in
transition probabilities (the Golden Rule).
Less familiar is the peculiar fact that here this
appears expressed as (\ref{nlnkApp}),
coming from an integral over momenta
of the form $\int dk/k$. This feature of
near horizon states was previously noticed in \cite{Parentani:1992me}
in the context of both the acceleration horizon of an Unruh detector,
and when counting the number of Hawking quanta in terms
of the free fall frequency they had near the horizon.
See
the discussion around
Eq. (3.44) in \cite{Brout:1995rd} for a brief account
of this.}

It is informative to compare our result with the mode density for
the static ``brick wall" reflecting boundary condition used
by 't Hooft in Ref.~\cite{'tHooft:1984re}. In that
model, the quantum field
is required to vanish at the wall, i.e. at a
fixed radius outside the
horizon.
't Hooft expressed the mode density
as the derivative of the number of radial nodes with respect to
frequency,  $\pi^{-1}\int dr\, (\partial
k_r/\partial\omega)$, where $\omega$ is the frequency with
respect to the Schwarzschild time coordinate.
Using the group velocity
relation $dr/dt=\partial\omega/\partial k_r$
this becomes simply $\pi^{-1}\int dt$. The
integral is the Schwarzschild time
lapse along the classical trajectory that rises from the wall and
stops at a turning point.

The one-bounce trajectory that rises and
then falls back to the wall is time-symmetric in terms of the
Schwarzschild time coordinate, so we may
extend the integral over $t$ to the full round trip integral
and divide by two.
This results in an expression that looks identical
to (\ref{nDtApp}), but it differs in that the time coordinate
is the Schwarzschild time rather than the
PG $\tau$
coordinate in (\ref{ds2}). However,
if we choose
$v(\infty) = 0, \k_v = \k$
then
these time coordinates
differ only by a function of radius, so
their lapses at any fixed radius are equal.
A small discrepancy remains because,
unlike t'Hooft's brick wall, our
regularization scheme leads to a
slightly asymmetric domain of integration.
The final point of the trajectory defining our mode density is at
horizon crossing on the inward branch, while the initial
point lies outside the horizon
where the momentum is equal to the cutoff.
The lapse of time associated with the extra
bit is of order $1/\Lambda$
however, and is thus negligible for our purposes.
Our cutoff mode density and that of the brick wall
prescription would thus agree if the cutoff momentum
occurred at the same location as the brick wall.
The comparison of the brick wall and momentum cutoffs
is discussed further in the text.

\subsection*{Free-fall vacuum: localization of particles and partners}

It follows from (\ref{KGDo}) that the modes
\beq
 \tphi^{+}_{\o\ell}(k) =\theta(k) k^{-i\o/\k-1}e^{-ik/\k}\chi_{\ell}(k),\label{phi+}
\eeq
have positive norm if  $\chi_\ell$ has a positive Wronskian.
These
are complete and orthogonal when $\o$ ranges over all values, both positive and negative. They thus
generate the Hilbert space of one-particle outgoing states.
The outgoing free-fall vacuum is defined by the condition that
it is annihilated by the
destruction operator
$ \hat a^{in}_{\o \ell}
= \la \tphi^{+}_{\o \ell}, \hat \phi\ra $
corresponding to these modes
because their FF frequency
$\Omega = \o - vk$ is positive, see (\ref{useful}).

The Fourier transforms of the modes (\ref{phi+}) have support on
both sides  of the horizon, even for $\ell=0$.
At the quantum level this gives rise to correlations
between field fluctuations on either side of the horizon in the
free-fall vacuum.
These correlations
are best
characterized using a different mode basis consisting of positive
Killing frequency ``particles" and negative Killing frequency
 ``partners", as we now explain.

The key mathematical ingredient is the decomposition
into terms analytic in the upper and lower half complex $k$-planes:
\beq
\theta(k) k^{-i\o-1} =
\frac{1}{e^{ \pi\o} - e^{- \pi\o}}\bigl[ e^{ \pi\o} (k-i \epsilon)^{-i\o-1}-
e^{- \pi\o} (k+i \epsilon)^{-i\o-1}\bigr]
\label{plusminus}\eeq
where $\epsilon \to 0^+$.
Here we have chosen the branch cut of the logarithm to run
along the negative $k$ axis.
(To simplify the notation we now adopt units with
$\k=1$.) The Fourier transform of the first term
 vanishes for  $x < 0$, since it is analytic in the
lower
half $k$-plane and bounded as $|k|\rightarrow\infty$.
Similarly the Fourier transform of
the second term
vanishes
for $x>0$.
This is not the whole story however, since the mode
(\ref{phi+})  contains the additional factor
$e^{-ik}\chi_\ell(k)$.\footnote{For zero angular momentum modes
$\chi_0=\exp(\pm i k/\k)$ the
sign of the Wronskian is $\pm$, and (\ref{useful}) becomes $(\o-\hv
k)\tu_{\o0}=\pm k \tu_{\o0}$. Those modes with positive free-fall frequency
and positive Wronskian therefore have only positive wave-vectors,
i.e. they are outgoing.
In the presence of small (large)
angular momentum the would-be purely outgoing modes
are partly (totally) reflected by the
angular momentum barrier
and fall back as ingoing modes. We can still identify these
as ``outgoing" however, in the sense that they have no incoming
part far from the horizon. These are the modes that are relevant
to the entanglement entropy.
Those with large $\ell$ constitute the dominant ones.}

Let us first consider
the localization in the relativistic case with
vanishing angular momentum.
In this case $\chi_{\ell=0}(k)\propto e^{\pm ik}$. The solution with positive Wronskian
is $e^{ik}$, which cancels the factor $e^{-ik}$ in the mode.
Thus the mode
splits into a sum of two
functions strictly localized inside and outside the
horizon. If we allow for angular momentum, then the solution $\chi_\ell(k)$
with positive Wronskian is no longer just $e^{ik}$, but is rather given by the
Hankel function,  $k^{1/2}H^{(1)}_\nu(k)$ in (\ref{Bessel}).
This introduces branch cuts
which lead to
the Fourier transform
of $(k-i\epsilon)^{-i\o-1}e^{-ik}k^{1/2}H^{(1)}_\nu(k)$
having support inside as well as outside the horizon.
The part inside corresponds to the reflected wave
(of low momentum $k \sim \k$)
that has bounced off the angular momentum barrier.

The addition of dispersion at a scale $\L$ further complicates the
localization properties
by modifying this time the UV part $k \sim \L \gg \k$ of the
waves.
As
shown  in \cite{Brout:1995rd,Balbinot:2006ua} the
story is qualitatively similar, except for a mixing of outgoing and
ingoing modes that corresponds to an extra pair creation process arising
from the
UV propagation of the modes.
The amplitude of this
is exponentially suppressed by a factor $\sim e^{-\pi\L/\kappa}$
in the case of quartic dispersion $F^2=k^2 + k^4/\Lambda^2$,
and can  therefore be safely neglected
for black holes with surface gravity
much less than $\Lambda$.

When (\ref{plusminus}) is inserted into the mode expression
(\ref{phi+})
the resulting two terms have KG
 norms with opposite signs.
This is best seen with the help of the identity
\beq (k\pm i\epsilon)^{-i\o-1}= \theta(k) k^{-i\o-1} -  \theta(-k)
e^{\pm\pi\o} (-k)^{-i\o-1}. \label{inverse} \eeq
The contribution from $\theta(k)$ has positive norm while
that from $\theta(-k)$ has negative norm, opposite
aside from the factor  $ e^{\pm\pi\o}$.
Hence
$(k- i\epsilon)^{-i\o-1}$   yields a
positive norm mode if $\o>0$, while
$(k+ i\epsilon)^{-i\o-1}$
yields a negative norm mode.
Taking into account the normalization
factor of each of these modes,
the annihilation
operator for (\ref{phi+})
with $\o > 0$ thus decomposes as
%
\beq
\frac{1}{\sqrt{1 - e^{- 2\pi \o}}}
\bigl[\hat b_{\o  \ell}  -
e^{- \pi\o} \hat c^\dagger_{ \o \ell}\bigr]\, .
\label{abc}\eeq
That is, it is a combination of  the
annihilator of
the $(k-i\epsilon)^{-i\o-1}$ mode
and the creation operator for the complex conjugate of
the $(k+i\epsilon)^{-i\o-1}$ mode,
which is a positive norm mode with Killing frequency $-\o$.
When $\o < 0$,
one finds (\ref{abc}) with
$b$ and $c$ interchanged, and $\o$ replaced by $-\o$.
The condition that the annihilator for (\ref{phi+}) (for all $\o$)
annihilates the state thus
becomes a statement that the state contains correlated pairs of
$b,c$
particles
with opposite 
Killing frequency.
Tracing over the negative 
Killing frequency
particles
yields a thermal state at the Hawking temperature
$T_H=\kappa/2\pi$. Thus, even in the dispersive
case, the near horizon free fall vacuum  has a
thermal character.

With vanishing angular momentum and no dispersion, these
correlated pairs are strictly localized,
with the positive Killing frequency one outside and the
negative Killing frequency one inside the horizon.
With angular momentum
added, the positive Killing frequency mode
falls back across the horizon with a low
radial
momentum.
With dispersion, there is a further blurring of the outside/inside distinction
which this time comes from the UV domain.
Nevertheless,
because $\L \gg \k$
there exists an intermediate domain $\L \gg k \gg \k$ wherein
(\ref{plusminus}) can
be used to define a decomposition into a
correlated pair of positive and negative Killing frequency
modes which are localized on either side of the horizon.
Moreover, when wavepackets are formed by integrating over $\o$, those
wavepackets will spend some time near the horizon behaving as do the ones with
no angular momentum and no dispersion.
The amount of time the one with positive $\o$ will spend
outside the horizon while its partner is inside is (by definition) given by
(\ref{nDtApp}).

Indeed,
in identifying the modes contributing to entanglement entropy, we use
the WKB description of these modes, and require that the positive Killing frequency
particle lies outside the horizon while the negative Killing frequency partner
lies inside. Having formed a wavepacket, at any given time the wavevector of the
particle and partner is given by the stationary point of the wavepacket integral
over $\o$. It follows from (\ref{plusminus}) and the fact that no other rapidly oscillating
function of $\o$ appears in the wavepacket that, at leading WKB order, this
wavevector is the {\it same} for the particle and partner. That common wavevector is then
used to locate the particle and partner using the dispersion relation with $\o>0$ for the
particle and $-\o$ for the partner.

\subsection*{Particle and partner trajectories}

In the near horizon region, the dispersion relation
(\ref{nhdisp})
reads
\beq
\o  = \k k x  - k +F(k,p;\L).
\label{eq1}
\eeq
Solving for the trajectory $x=x(k,\o,p; \L)$ we have
\beq
x(\o,p,k; \L) = x_0(k,p;\L) + \o/\k k,
\label{x}
\eeq
with
\beq
x_0(k,p;\L)= ( k-F(k,p;\L))/\k k.
\label{xmean}
\eeq
As just explained, the particle and partner have the
same value of $k$ at any given time,
and opposite values of $\o$. Hence the mean position
(\ref{xmean}) of the pair is independent of $\o$,
while the separation of the particle and partner
is universally given by
\beq
\D x = 2 \o /\k k,
\eeq
independent of $p$ and the form of the dispersion.
This reflects at the classical level the fact, seen in
(\ref{factoreddisp}),  that the
modes always split into a factor independent of
$p$ and $F$ and a factor independent of $\o$.
This result,
as many others,
is identical to what was found
in the 1+1 case studied in \cite{Balbinot:2006ua}
because the transverse
momentum $p$
{only} enters as a modification of
the dispersion function $F$.

The space time trajectories $x(\tau)$ are determined by
(\ref{x}) since $k(\tau)$ is
known. As seen in
(\ref{kdot}), the $k$-trajectories are simply
given by $k(\tau) = k_0 \exp(-\k \tau)$,
again independently of the value of $p$
and the dispersion relation governed by $F$.

\end{document}